\documentstyle[11pt,epsfig]{article}
\begin{document}
\newcommand{\be}{\begin{equation}}
\newcommand{\ee}{\end{equation}}
\newcommand{\ba}{\begin{eqnarray}}
\newcommand{\ea}{\end{eqnarray}}
\newcommand{\no}{\nonumber \\}
\newcommand{\gsim}{\mathrel{\hbox{\rlap{\lower.55ex \hbox {$\sim$}}
                   \kern-.3em \raise.4ex \hbox{$>$}}}}
\newcommand{\lsim}{\mathrel{\hbox{\rlap{\lower.55ex \hbox {$\sim$}}
                   \kern-.3em \raise.4ex \hbox{$<$}}}}
\def\be{\begin{eqnarray}}
\def\ee{\end{eqnarray}}
\def\bea{\be}
\def\eea{\ee}
\newcommand{\e}{{\mbox{e}}}
\def\del{\partial}
\def\vr{{\vec r}}
\def\vk{{\vec k}}
\def\vq{{\vec q}}
\def\vp{{\vec p}}
\def\vP{{\vec P}}
\def\vt{{\vec \tau}}
\def\vs{{\vec \sigma}}
\def\vJ{{\vec J}}
\def\vB{{\vec B}}
\def\hatr{{\hat r}}
\def\hatk{{\hat k}}
\def\roughly#1{\mathrel{\raise.3ex\hbox{$#1$\kern-.75em%
\lower1ex\hbox{$\sim$}}}}
\def\lsim{\roughly<}
\def\gsim{\roughly>}
\def\fm{{\mbox{fm}}}
\def\vx{{\vec x}}
\def\EM{{\rm EM}}
\def\barp{{\bar p}}
\def\zz{{z \bar z}}
\def\mus{{\cal M}_s}
\def\abs#1{{\left| #1 \right|}}
\def\ve{{\vec \epsilon}}
\def\nlo#1{{\mbox{N$^{#1}$LO}}}
\def\MS{{\mbox{M1V}}}
\def\mut{{\mbox{M1S}}}
\def\Qt{{\mbox{E2S}}}
\def\rM{{\cal R}_{\rm M1}}\def\rE{{\cal R}_{\rm E2}}
\def\la{{\Big<}}
\def\ra{{\Big>}}

\def\J#1#2#3#4{ {#1} {\bf #2} (#4) {#3}. }
\def\PRL{Phys. Rev. Lett.}
\def\PL{Phys. Lett.}
\def\PLB{Phys. Lett. B}
\def\NP{Nucl. Phys.}
\def\NPA{Nucl. Phys. A}
\def\NPB{Nucl. Phys. B}
\def\PR{Phys. Rev.}
\def\PRC{Phys. Rev. C}

\renewcommand{\thefootnote}{\arabic{footnote}}
\setcounter{footnote}{0}

\vskip 0.4cm
\hfill {\bf KIAS-P99045}

\hfill {\today}
\vskip 1cm

\begin{center}
{\LARGE\bf Elastic Parton-Parton Scattering\\
 From  AdS/CFT }

\date{\today}

\vskip 1cm
{\large
Mannque Rho$^{a,b}$\footnote{E-mail: rho@spht.saclay.cea.fr},
Sang-Jin Sin$^{a,c}$\footnote{E-mail: sjs@hepth.hanyang.ac.kr} and
Ismail  Zahed$^{a,d}$\footnote{E-mail: zahed@nuclear.physics.sunysb.edu}
}



\end{center}

\vskip 0.5cm

\begin{center}

$^a$
{\it School of Physics, Korea Institute for Advanced Study,
Seoul 130-012, Korea}

$^b$
{\it Service de Physique Th\'eorique, CE Saclay,
91191 Gif-sur-Yvette, France}

$^c$
{\it Department of Physics, Hanyang University, Seoul 133-791, Korea}

$^d$
{\it Department of Physics and Astronomy,
SUNY-Stony-Brook, NY 11794}

\end{center}

\vskip 0.5cm

\begin{abstract}
Using the AdS/CFT correspondence and the eikonal approximation,
we evaluate the elastic parton-parton
scattering amplitude at large $N$ and strong coupling $g_{YM}^2N$ in N=4 SYM.
We obtain a scattering amplitude that reggeizes and 
unitarizes at large $\sqrt{s}$. 
\end{abstract}

\newpage

\renewcommand{\thefootnote}{\#\arabic{footnote}}
\setcounter{footnote}{0}


{\bf 1.\,\,} Elastic quark-quark and gluon-gluon scattering at
large $s$ and fixed $t$ (Mandelstam variables)
pertains to the domain of non-perturbative
QCD. Theoretical procedures based on resuming large classes of
perturbative contributions have been proposed~\cite{BFKL,POMERON},
partially accounting for the reggeized form of the scattering
amplitude and the phenomenological success of pomeron/odderon
exchange models~\cite{POMERON} (and references therein). In an
inspiring approach, Nachtmann~\cite{NACHTMANN} and
others~\cite{VERLINDE,OTHERS,arefeva} suggested to use
non-perturbative techniques for the elastic scattering amplitude.
In the eikonal approximation and to leading order in $-t/s$ the
quark-quark amplitude at large $s$ was reduced to a correlation
function of two light-like Wilson-lines. The latter was assessed in
the stochastic vacuum model \cite{NACHTMANN}.

Recently, Maldacena~\cite{MALDACENA} has made the remarkable
conjecture that the large $N$ behavior of $N=4$ supersymmetric
gauge theory is dual to the string theory in a non-trivial
geometry. This AdS/CFT conjecture, made more precise by Gubser, Klebanov
and Polyakov~\cite{gkp} and by Witten~\cite{WITTEN} and extended to
the non-supersymmetric case by Witten~\cite{thermal}, provides an
interesting and nonperturbative avenue
for studying gauge theories at large $N$ and strong coupling
$g_{YM}^2N$. In particular the heavy quark potential was found to be
Coulombic for supersymmetric theories and linear for
non-supersymmetric theories~\cite{WITTEN,yanki}.

In this letter we suggest to use the AdS/CFT approach to analyze the
elastic parton-parton scattering amplitude in the eikonal approximation
for $N=4$ SYM. In section 2 we briefly discuss the salient features of
 the parton-parton reduced elastic amplitude at large $\sqrt{s}$.
The scattering amplitude in the eikonal approximation is reduced to the
Fourier transform of the connected part of a correlator of
two time-like Wilson lines.
In section 3 we use the AdS/CFT approach to calculate the correlator.
Following \cite{MALDA2,REY},
we propose that it is given by the minimum (regularized) area of the
world-sheet with the time-like parton trajectories at its boundaries,
and give a simple variational method
that allows for a closed-form expression for the minimal area
in the presence of a finite time cutoff.
In section 4 we summarize and conclude.

\vskip 1cm

{\bf 2.\,\,}
First consider the quark propagator in an external non-Abelian gauge field.
In the first quantized theory, the propagator from $x$ to $y$ in Minkowski
space reads
\be
{\bf S}(x,y;A) = \la x | \frac i{i\rlap/\nabla -m+i0} | y\ra =
\int_0^{\infty} dT e^{-i(m-i0)T} \la x | e^{-\rlap/\nabla\, T} |y\ra
\label{2}
\ee
that is~\cite{SPINFACTOR,POLYAKOV}
\be
{\bf S}(x,y;A) =
\int_0^{\infty} \, dT \,
e^{-i(m-i0) T} \,\int_x^y d[x] \, \,\delta (1-\dot{x}^2)\,\,
{\bf P}_ce^{ig_{YM}\int ds A\cdot \dot{x}}\,\,
\frac 12 {\bf P}_s e^{-\frac i2 \int ds\, \sigma^{\mu\nu}\dot{x}_{\mu}
\dot{\dot{x}}_{\nu}}
\label{2x}
\ee
where ${\bf P}_{c,s}$ are orderings over color and spin matrices,
and $\sigma_{\mu\nu}=[\gamma_{\mu},\gamma_{\nu}]/2$. The first
exponent in the path integral
is an arbitrary Wilson line in the fundamental representation of
SU(N)$_c$, and the second exponent is a string of infinitesimal
Thomas-precession along the Wilson line.
The integration is over all paths with $x_{\mu}(0)=x_{\mu}$ and
$x_{\mu}(T)=y_{\mu}$,
where $T$ is the proper-time~\cite{POLYAKOV}. The dominant
contributions come from those paths with $T\leq 1/m$.
Heavy quarks travel shorter in
proper-time than light quarks and the mass gives an effective
cutoff of the proper-time range.  This observation will be
important below.

A quark with large momentum $p$ travels on a straight line with 4-velocity
$\dot{x}_{\mu}=u_{\mu}=p_{\mu}/m$  and $u^2=1$.
Throughout, we will distinguish between the
4-velocity $u_{\mu}$ and the instantaneous
3-velocity ${v}=dx/dt=p/E\leq 1$.
For a straight trajectory, the 4-acceleration $\dot{\dot{x}}_{\mu}=a_{\mu}=0$ and the
spin factor drops. This is the eikonal approximation for ${\bf S}$ in which an
ordinary quark transmutes to a scalar quark. The present argument applies to
any charged particle in a background gluon field, irrespective of its spin
or helicity. The only amendments are: for antiquarks the 4-velocity $v_{\mu}$ is
reversed in the Wilson line and the color matrices are in the complex
representation, while for gluons the Wilson lines are in the
adjoint representation. With this in mind, quark-quark scattering can be
also extended to quark-antiquark, gluon-gluon or scalar-scalar scattering.
We note that for quark-antiquark scattering the elastic
amplitude dominates at large $\sqrt{s}$ since the annihilation
part is down by $\sqrt{-t/s}$.

Generically, we will refer to elastic parton-parton scattering as
\be
Q_A(p_1) + Q_B(p_2) \rightarrow Q_C(k_1) +Q_D(k_2)
\label{3}
\ee
with $s=(p_1+p_2)^2$, $t=(p_1-k_1)^2$, $s+t+u=4m^2$. We denote by
$AB$ and $CD$ respectively, the incoming and outgoing
color and spin of the quarks (polarization for gluons). Using
the eikonal form for (\ref{2}) and LSZ reduction,
the scattering amplitude ${\cal T}$
may be reduced to~\cite{NACHTMANN,VERLINDE,OTHERS}
\be
{\cal T}_{AB,CD} (s,t) \approx
-2is \int d^2b\,\, e^{iq_{\perp}\cdot b}
\la \left({\bf W}_1 (b) \right)_{AC}
\left( {\bf W}_2 (0) \right)_{BD}\ra_c
\label{4}
\ee
where
\be
{\bf W}_{1,2}(z)= {\bf P}_c {\rm
exp}\left(ig_{YM}\int_{-\infty}^{+\infty}\,d\tau\, A(b+u_{1,2}\tau)\cdot
u_{1,2}\right).
\label{5}
\ee
We are using the normalization $\left<{\bf W}\right> =1$, and
retaining only the connected part in (\ref{4}). The normalization
can be relaxed if needed. The 2-dimensional
integral in (\ref{4}) is over the impact parameter $b$ with
$t=-q_{\perp}^2$. In the CM frame $p_1\approx (E,E,0_{\perp})$,
$p_2\approx (E,-E,0_{\perp})$, $q=p_1-k_1\approx (0,0,q_{\perp})$
and $s\approx 4E^2$. The averaging in (\ref{4}) is over the gauge
configurations using the QCD action. The total cross section for
$\sqrt{s}>>\sqrt{-t}>0$ follows from (\ref{4}) in the form $\sigma
= {Im\,\,{\cal T}}/s < {\rm ln}^2 s$, where the last inequality is
just the Froissart bound.

The amplitude (\ref{4}) allows for two gauge-invariant decompositions
(repeated indices are summed over)
\be
{\cal T}_1={\cal T}_{AB,AB}\qquad\qquad
{\cal T}_2={\cal T}_{AB,BA}
\label{8}
\ee
assuming that the gluon-gauge fields are periodic
at the end-points. 
We note that ${\cal T}_2$ is down by $1/N$ in comparison
to ${\cal T}_1$. For gluon-gluon scattering the lines
are doubled in color
space (i.e., adjoint representation) and several gauge-invariant
contractions are possible. For quark-quark scattering the singlet
exchange in t-channel is 0$^+$ (pomeron) while for quark-antiquark
it is 0$^-$ (odderon) as the two differ by charge conjugation.

\vskip 1cm

{\bf 3.\,\,}
In the eikonal approximation the parton-parton scattering amplitude
is related to an appropriate correlator of two Wilson lines. The
typical duration of these light-like lines is $T\sim 1/m$
as we noted above. We now suggest to analyze the gauge-invariant
correlators using the AdS/CFT
\cite{MALDACENA,WITTEN} approach for N=4 SYM.

The correlation function in large
$N$ and strong coupling $g_{YM}^2N$  can be obtained from the
minimal surface in the five-dimensional AdS space,
\cite{MALDACENA,REY,WITTEN} with the light-like
Wilson lines at its boundaries as shown in Fig. 1.
\begin{figure}[t]
\centerline{\epsfig{file=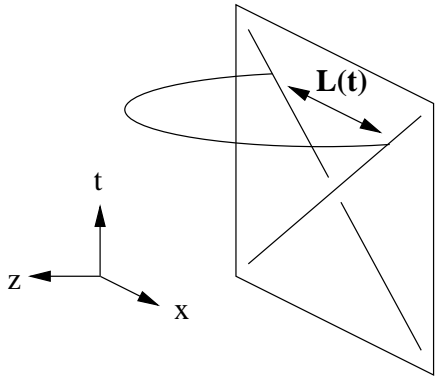,height=5cm}}
\caption{Wilson lines for moving partons with the attached AdS surface.}
\label{fig1}
\end{figure}
The classical action for a string world-sheet is
\be
S=-\frac{1}{2\pi \alpha'}\int d\xi d\sigma
\sqrt{\det(G_{MN}\partial_\alpha X^M\partial_\beta X^N)}. \label{action}
\ee
The AdS metric  $G_{MN}$ in Poincar\'e coordinates is given by
\be
ds^2 =R^2 \frac{-dt^2+dx^2+dy^2+dw^2 +dz^2}{z^2} \label{metric},
\ee
where $R=(2 g^2_{YM} N\alpha'^2)^{1/4}$ is the radius of the AdS
space, with $2\pi g_{st}=g_{YM}^2$. The AdS space has a boundary in
Minkowski space $M_4$ at $z=0$. The boundary condition on the
string world-sheet is given by the two time-like trajectories
\be
x={v}t={u}\tau, y=0, z=0; \mbox{ and } \; x=-{v}t, y=b, z=0
\label{bc}
\ee
with 3-velocities ${v}={u}/\gamma$ and $t$ the real time
\footnote{We are using $t$ alternatively for the Mandelstam variable
and the real time, and $u$ alternatively for the Mandelstam variable and
the 4-velocity. In each case the meaning should be clear from the text.}.
The minimal surface associated to (\ref{action}-\ref{bc})
leads to a set of coupled partial differential equations,
which we have not yet managed to solve exactly.
Instead, we provide a variational estimate as we now explain.

First, we divide the string world-sheet by constant time slices,
each containing a string connected to two boundary points.
Then we assume that for the minimal surface, this string length is minimal.
So by finding the minimal length and carrying the integration over time, 
we will obtain
an approximate minimal area. Specifically, we choose an orthogonal
coordinate system $(\xi,\sigma)$ with the property $\partial_\xi
X^\mu \partial_\sigma X^\mu=0$  on the world-sheet. Then,
\be
S= -\frac{R^2}{2\pi \alpha'} \int d\xi d\sigma \sqrt{(\partial_\xi X^\mu)^2}
\sqrt{\frac{(\partial_\sigma X^\mu)^2}{z^4}}.
\ee
Let
\be
l(t) := \int_{L(t)} d\sigma
\sqrt{\frac{(\partial_\sigma x)^2+(\partial_\sigma y)^2+(\partial_\sigma w)^2 +(\partial_\sigma z)^2}{z^4}}
\ee
be the `length' of the string ending on the two receding
quarks at the boundary
with separation $L(t)$.
$l(t)$ depends on $t$ only through $L(t) (=L(\tau))$, so that
$l(t)$ and $l(\tau)$ represent the same quantity.
First, we minimize this `length' and then form the area by adding up
the areas of the strips between the hyper-planes of time $t$ and $t+dt$.
The height of the strip at the boundary is
$dt\sqrt{1-{v}^2}=d\tau$ where $\tau$ is the proper-time at the boundary.
The height of the strip at the central point of the
string is given by $dt\sqrt{1-{\dot {z_0}(t)}^2}$
where ${z_0}(t)$ is the maximum value of $z$ at fixed $t$.
Therefore by the trapezoidal rule, the minimal area is given by
\be
A_{min} \simeq \frac{1}{2}\int dt\,
\left(\sqrt{1-v^2}+\sqrt{1-{\dot {z_0}}^2}\right)\,
l_{min}(t)\label{trapezoid},
\ee
where $l_{min}(t)$ is the minimal
length for a fixed time slice. To
summarize: we have replaced $d\xi
\sqrt{(\partial_\xi X^\mu)^2}$ by  $dt\, (\sqrt{1-v^2}
+\sqrt{1-{\dot {z_0}}^2})/2$, which is $\sigma$ independent. 
The latter substitution is made in the CM frame.
Lorentz invariance follows
by rewriting the results in terms of Lorentz scalars.  

The minimal length 
$l_{\min}(t)$ can be found by choosing a coordinate system such
that the two quarks are located at $x=L(\tau)/2$ 
and $x=-L(\tau)/2$.
Then
\be
l(t) =\int_{-L(t)/2}^{+L(t)/2} dx \sqrt{ \frac{1+z'^2}{z^4}} \label{sec},
\ee
where $L(t)=\sqrt{b^2+4v^2t^2}$ (${u}\tau={v}t$)
is the separation between the
two time-like receding partons at the boundary $z=0$.
In the instantaneous approximation, 
the string adjusts instantaneously
to the change in the minimal length $L(t)$ at the boundary.~\footnote{
This implicit approximation follows from the fact that we have neglected
the orthogonality constraint imposed on the coordinate system
in the variational estimate. Its physical consequence will be addressed
below.} It follows that 
the problem of finding a minimal $l(t)$ is almost identical to the
problem of finding a static $q{\bar q}$ potential~\cite{MALDA2,REY}.
The result for a properly regularized length is
\be
l(t)= -\frac{c_0}{L(t)}
\ee
with $c_0=(2\pi)^{3}/\Gamma(\frac{1}{4})^4$ and ${z_0}(t)$ given by
\be
{z_0} (t) = c_1 L(t) ,
\ee
where $c_1=1/\sqrt{c_0}\simeq 0.834$.
Hence under the instantaneous approximation, the `area' is obtained by 
integrating the static potential.
\ba
A_{min} &\simeq&-\frac{1}{2}\int_{-T}^{T} dt\, \left(\sqrt{1-v^2}
+\sqrt{1-{\dot{z_0}}^2}\right)
\,\frac{c_0}{\sqrt{b^2+4{v}^2 t^2}}\no
&=& -\left(\sqrt{1-v^2}+\sqrt{1-(2c_1{v})^2}\right)
\frac{c_0}{2{v}}\sinh^{-1} \left(\frac{2{v}T}{b}\right).
\label{15}
\ea
According to the AdS/CFT correspondence~\cite{WITTEN,MALDA2},
the connected part of the Wilson-line
correlator in N=4 SYM at large $N$ and fixed $g_{YM}^2N$ is
\be 
\la{\bf WW}\ra_c  \approx
\exp(iS)= 
\exp\left[ic_0 \sqrt{2 g_{YM}^2N}\left(\sqrt{1-v^2}+ \sqrt{1-(2c_1{v})^2}\right)
\frac{1}{2{v}} \ln \left(\frac{4{v}T}{b}\right)\right].
\label{16}
\ea
This result may be
contrasted with the one-gluon exchange contribution to the
connected and untraced correlator (dropping color factors)
\be
\la {\bf WW}\ra_{1c} \approx - \frac{g_{YM}^2}{4\pi^2}\, 
\int_{-\infty}^{+\infty} d\tau_1 d\tau_2
\frac {u_1\cdot u_2}{\left(- (u_1\tau_1 -u_2\tau_2)^2 + b^2 \right)}
= i\frac{g_{YM}^2}{4\pi} \left(v+\frac 1v\right)
\,{\rm ln} \left(\frac{T}{ b}\right).
\label{18}
\ee
The natural infrared cutoff in the problem is  the mass; $T=1/\mu\sim 1/m$.
For quarks and gluons, it is simply their constituent mass.
\footnote{For parallel moving quarks, the result is $1/b$ instead of 
${\rm ln}\,b$. Coulomb's law is 2-dimensional for non-parallel moving
light-like quarks
and 3-dimensional for heavy or parallel moving light-like quarks.}
In QED, (\ref{18}) exponentiates with a noticeable difference from
(\ref{16}) : the time dilatation factors generated by
the string are absent in QED. This very difference will cause the 
scattering amplitude to
{\it reggeize} in our case instead of eikonalize as we will show below.

To restore Lorentz invariance, we can rewrite eq.(\ref{16}) 
in terms of the Mandelstam variable $s$,
\be
\la {\bf WW}\ra_c &\approx&
\left(\frac{4\sqrt{1-4m^2/s}}{\mu b}\right)^{ic_0\sqrt{2 g_{YM}^2N}
\left(\frac{m}{\sqrt{s-4m^2}}+
\frac{1}{2}\sqrt{\frac{s}{s-4m^2}-(2c_1)^2}\right)}
\label{correct}
\ee
where we have made the substitutions :
${u}=\sqrt{\frac{s}{4m^2}-1}$, ${v}=\sqrt{1-\frac{4m^2}{s}}$ and
$T=1/\mu$ for the time cutoff. At this stage,
several remarks are in order. 

\begin{itemize}
\item Notice that $\sqrt{1-(2c_1{v})^2}$ is purely imaginary 
for ${v}>1/2c_1\simeq 0.6$, resulting in a suppression 
of the scattering amplitude at large $\sqrt{s}$.
This happens because the central part of the string
moves into the AdS space with a velocity $\sim 2c_1v >1$,
which is classically forbidden by the 5 dimensional kinematics.
This is a flaw of the instantaneous approximation we discussed
above, that can be ultimately resolved by an analytical or numerical 
investigation of the exact solution to the minimal surface problem. 
Here, we observe that this pathology can be removed by physical 
arguments on the CFT side. 
Indeed, from (\ref{correct}) we note that the pathological 
behavior of the string yields a new branch-point singularity in 
the s-channel, besides the expected free threshold  at
$s=4m^2$. In a non-confining and conformally invariant SYM theory, 
this is unphysical. This singularity disappears if and only if 
$2c_1$ is renormalized to 1. This means that the exact treatment 
of the minimal surface should yield a result where the maximum
speed of the central point of the string is 1, in accordance with
relativity. When this happens,
\be
\la {\bf WW}\ra_c &\approx&
\left(\frac{4\sqrt{1-4m^2/s}}{m\mu b}\right)^{ic_0\sqrt{2 g_{YM}^2N}
\left(\frac{2m}{\sqrt{s-4m^2}}\right)}
\label{final}
\ee

\item Eq.(\ref{15}) might be interpreted 
to suggest using the proper-time 
$d\tau=\sqrt{1-v^2}dt$ instead of the global varaible $t$ 
in the final stage of the calculation.
It is important to realize that if the $fixed$
time cutoff $T=1/{\mu}$ was substituted by a $fixed$
proper time cutoff then the result would be (\ref{final})
with the substitution $1/\mu\rightarrow \sqrt{s}/2m\mu$. 
The former ($t$) is favored by the string theory calculation of the minimal 
surface and to compare the well-known eikonal form
in the Abelian case such as 
QED~\cite{OTHERS,cheng} as well as 
the result of the perturbative calculation eq.(\ref{18}). 
The latter($\tau$) is favored by manifest Lorentz invariance and the 
representation (\ref{2}-\ref{2x}). The situation is such that
string theory calculation with time cutoff gives the gaugy theory 
results with formalism using the proper-time.
Fortunately, whether we use the proper-time or time cutoff,
both lead to the same scattering amplitude asymptotically
(see below). Note that the elastic amplitude 
involves typically momenta of order $\sqrt{-t}$, so that
$b\sim 1/\sqrt{-t}$. As a result, the `cross singularity' of the two
Wilson loops at $b=0$~\cite{OTHERS}  is dynamically regulated for 
fixed Mandelstam variable $t$.

\item The Minkowski AdS/CFT approach followed here is subtle. For example, 
the concept of a minimum surface is not well defined in a metric with
indefinite signature. A more rigorous treatment is 
to setup the problem in a metric with 
an Euclidean signature and then perform a  Wick-rotation of the outcome.
We have checked that our present answer is unchanged
by this procedure. The factor $i$ in the exponent of (\ref{final})
follows from $1/v_E$ ($v_E={dx}/{d\,t_E}, it=t_E$),
which is real with an Euclidean signature but imaginary with a Minkowski 
signature.   
\end{itemize}

Using (\ref{final}), the gauge-invariant combination of the
parton-parton scattering amplitude (\ref{8}) now reads
\be
{ \cal T}(s,t)\approx  4\pi 
\alpha(s)\frac{\Gamma(1-i\alpha(s))}{\Gamma(1+i\alpha(s))} 
\left(\frac{2s}{-t}\right)\, \left(\frac{2\sqrt{-t}}{\mu}
\right)^{2i\alpha(s)}
\label{amp}
\ee
for large $N$ and fixed $g_{YM}^2N$. 
Here the gamma functions come from $2\pi \int_0^\infty d
b' b'^{1-2i\alpha(s)}J_0(b')$ with $\alpha (s)$ given by
\be
\alpha(s)={c_0\sqrt{2 g_{YM}^2N}\left[ \frac{m}{\sqrt{s-4m^2}} \right]}.
\ee
A similar behavior follows from a proper time cutoff through the
substitution $1/\mu\rightarrow \sqrt{s}/2m\mu$.
The result (\ref{amp})
is reminiscent of the QED result~\cite{cheng},
but with important differences. The amplitude 
has a nonperturbative dependence on $g_{YM}^2N$,
much like the static potential~\cite{MALDA2,REY}.
The amplitude reggeizes and unitarizes at large $\sqrt{s}$. 
Indeed, asymptotically ($s >> -t,m^2$)
\be
{\cal T} (s,t) \approx 4\pi c_0\sqrt{2g^2_{YM} N}
\left(\frac {2s}{-t}\right)\,\left(\frac s{m^2}\right)^{-0.5}
\label{REGGE}
\ee
which is real (no inelasticity), independent of the cutoff $\mu$, 
and with a negative
intercept of $-0.5$. The zero imaginary part 
and the nonzero intercept  are both tied to the 
the occurrence of a string in the AdS space
as is explicit from (\ref{16}). On the boundary, the receding partons with 
momenta $\sqrt{s}$ define a range in rapidity
space of the order of ${\rm ln}\,s$. Powers of ${\rm ln}\,s$
count the number of `gluons' exchanged in the t-channel. Since (\ref{REGGE})
can be written as a power series in ${\rm ln}\,s$, it
contains terms with an infinite number of gluon exchange.
A similar observation was also made by
Verlinde and Verlinde~\cite{VERLINDE} in the process of
mapping high-energy scattering onto a two-dimensional sigma model.

Finally, the present arguments also show that at large $N$ and strong $g_{YM}^2N$,
the cross sections for quark-quark and quark-antiquark scattering
are the same. The gluon-gluon
scattering amplitude could be calculated
similarly with the substitution $N\rightarrow N^2$,
due to the adjoint representation of the gluon.

\vskip 1.5cm
{\bf 4.\,\,} We have presented arguments for evaluating the elastic
parton-parton scattering amplitude at large $N$ and strong $g_{YM}^2N$
in N=4 SYM. Although the latter is conformally invariant in $M_4$
the appearance of the string picture and the necessity to regulate
the elastic contribution in time has
led to a reggeized behavior that unitarizes at large $\sqrt{s}$.
The result cannot be reached by perturbation theory. The nature
of the result depends sensitively on the string character of the
underlying description and hence is not applicable to Abelian-like
theories such as QED.

Our main result follows from a physically motivated variational estimate
of the minimal surface in the AdS space. The exact form of the extremal
surface is too involved to be written down analytically.
Although an exact result would of course be ideal, we do not expect our 
estimate of the parton-parton cross section to change appreciably.
Indeed, the dominant contributions arise from scattering with large impact 
parameter $b\sim 1/\sqrt{-t}$, for which our approximation for the extremal
surface should be legitimate.
For a large impact parameter $b$, the extremal surface is smoothly twisted
and the eikonal approximation is also good.

Finally and from a different standpoint,
Verlinde and Verlinde~\cite{VERLINDE}  have shown that at large $\sqrt{s}$ the
elastic amplitudes in QCD follows from a two-dimensional sigma model with 
conformal symmetry, where the latter is broken when the light-like quark 
lines are regulated in the time-like direction. Does the large $N$ 
effective action derived by Verlinde and Verlinde~\cite{VERLINDE}
map onto an AdS type action? Is there a way to do better than the variational 
estimate made in the present analysis? How could we recover the true Regge 
behavior of non-supersymmetric QCD?
Some of these questions will be addressed in a forthcoming publication.

\vskip 2.5cm

{\bf Acknowledgments}
\vskip .35cm
We wish to thank KIAS for their generous hospitality and support during this
work. We are grateful to Kyungsik Kang, Taekoon Lee, Maciek Nowak and Martin
Rocek for discussions, and Romuald Janik for comments on the manuscript. We
are especially thankful to Igor Klebanov for comments and helpful suggestions.
The work of IZ was supported in part by US-DOE grant DE-FG-88ER40388 and that
of SJS by the program BSRI-98-2441.

\vskip 1.5cm

{\it Note Added}

\vskip .35cm
After posting our paper, \cite{JANIK} appeared in
which a similar approach is suggested for scattering
between colorless states at large impact parameters.


\end{document}